\documentclass[fleqn,usenatbib]{mnras}

\usepackage{newtxtext,newtxmath}
\usepackage[T1]{fontenc}

\DeclareRobustCommand{\VAN}[3]{#2}
\let\VANthebibliography\thebibliography
\def\thebibliography{\DeclareRobustCommand{\VAN}[3]{##3}\VANthebibliography}

\usepackage{graphicx}	% Including figure files
\usepackage{amsmath}	% Advanced maths commands
\title[Radio observations of AT2020opy]{Radio observations of the tidal disruption event AT2020opy: a luminous non-relativistic outflow encountering a dense circumnuclear medium}

\author[A. J. Goodwin et al.]{
A. J. Goodwin,$^{1}$\thanks{E-mail:ajgoodwin.astro@gmail.com}, J. C. A. Miller-Jones$^{1}$, S. van Velzen$^{2}$, M. Bietenholz$^{3,4}$, J. Greenland$^{1}$,
\newauthor
B. Cenko$^{5}$,
S. Gezari,$^{6}$,
A. Horesh$^{7}$, 
G. R. Sivakoff$^{8}$, 
L. Yan$^{9}$,
W. Yu$^{10}$,
and X. Zhang$^{10, 11}$\\
$^{1}$International Centre for Radio Astronomy Research -- Curtin University, GPO Box U1987, Perth, WA 6845, Australia\\
$^{2}$Leiden Observatory, Leiden University, PO Box 9513, 2300 RA Leiden, The Netherlands \\
$^{3}$ Department of Physics and Astronomy, York University, Toronto, M3J 1P3, Ontario, Canada\\
$^{4}$ SARAO/Hartebeesthoek Radio Observatory, PO Box 443, Krugersdorp, 1740, South Africa \\
$^{5}$ Astrophysics Science Division, NASA Goddard Space Flight Center, Mail Code 661, Greenbelt, MD 20771, USA\\
$^{6}$ Space Telescope Science Institute, 3700 San Martin Drive, Baltimore, MD 21218, USA \\
$^{7}$ Racah Institute of Physics, The Hebrew University of Jerusalem, Jerusalem 91904, Israel \\
$^{8}$ CCIS 4-181 University of Alberta, Edmonton Alberta T6G 2E1, Canada  \\
$^{9}$ Caltech Optical Observatories, California Institute of Technology, Pasadena, CA 91125, USA\\
$^{10}$ Shanghai Astronomical Observatory, Chinese Academy of Sciences, 80 Nandan Road, Shanghai 200030, China \\
$^{11}$ University of Chinese Academy of Sciences, 19A Yuquanlu, Beijing 100049, China \\
}

\date{Accepted XXX. Received YYY; in original form ZZZ}

\pubyear{2022}

\begin{document}
\label{firstpage}
\pagerange{\pageref{firstpage}--\pageref{lastpage}}
\maketitle

\begin{abstract}
% Introduction
Tidal disruption events (TDEs) occur when a star passes too close to a supermassive black hole and is destroyed by tidal gravitational forces. Radio observations of TDEs trace synchrotron emission from outflowing material that may be ejected from the inner regions of the accretion flow around the SMBH or by the tidal debris stream. 
%The `gap' or problem
Radio detections of tidal disruption events are rare, but provide crucial information about the launching of jets and outflows from supermassive black holes and the circumnuclear environment in galaxies.
%To date, there are radio detections of $<20$ published TDEs. 
%"Here we show.."
Here we present the radio detection of the TDE AT2020opy, including three epochs of radio observations taken with the Karl G. Jansky's Very Large Array (VLA), MeerKAT, and upgraded Giant Metrewave Radio telescope.
%The overall approach, key results, and conclusions
AT2020opy is the most distant thermal TDE with radio emission reported to date, and from modelling the evolving synchrotron spectra we deduce that the host galaxy has a more dense circumnuclear medium than other thermal TDEs detected in the radio band. Based on an equipartition analysis of the synchrotron spectral properties of the event, we conclude that the radio-emitting outflow was likely launched approximately at the time of, or just after, the initial optical flare. We find no evidence for relativistic motion of the outflow. 
%The advance over previous work, the implications
The high luminosity of this event supports that a dense circumnuclear medium of the host galaxy produces brighter radio emission that rises to a peak more quickly than in galaxies with lower central densities. 

\end{abstract}


%%%%%%%%%%%%%%%%%%%%%%%%%%%%%%%%%%%%%%%%%%%%%%%%%%

%%%%%%%%%%%%%%%%% BODY OF PAPER %%%%%%%%%%%%%%%%%%

\section{Introduction}

% - opening
%     - Introduce general topic, context
%     - importance of research area
%     - background information
%     - current research focus
When a star passes within the tidal radius of a black hole, the star can be destroyed, producing a bright flare of electromagnetic radiation visible from radio to X-ray wavelengths \citep[e.g.][]{Rees1988}. 
The afterglow emission from such tidal disruption events (TDEs) at different wavelengths gives insight into the process by which the star was destroyed, the formation of accretion disks around black holes, the magnetic and gravitational fields of the central black hole, and the nuclear environment of the host galaxy \citep[e.g.][]{Lodato2011}. 

TDEs show diverse optical, X-ray, and radio properties, thought to be explained by the circumstances surrounding the stellar disruption and subsequent behaviour of the debris such as the SMBH mass, viewing angle, impact parameter as well as the circumnuclear environment of the host galaxy. Simulations have shown that the bound stellar debris may circularise to form an accretion disk \citep[e.g.][]{Bonnerot2016,Hayasaki2016,Liptai2019,Bonnerot2020,Mummery2020}, emitting X-ray radiation from accretion onto the SMBH, and optical radiation, from either re-processing of the X-rays in the accretion disk or stream-stream collisions of the tidal debris %\citep[e.g.][]{Strubbe2009,Auchettl2017,vanVelzen2020,Gezari2021}. 
\citep[e.g. see][for a review]{Roth2020}.
The time taken for the stellar debris to circularise and accretion to begin onto the SMBH is a matter of debate, with physical system properties such as the SMBH mass, stellar orbit, and stellar properties thought to affect the organisation of the debris \citep{Hayasaki2016,Liptai2019,Lu2019}. Observationally, X-ray properties of TDEs are extremely diverse \citep{Auchettl2017}, with some events never detected in X-rays, others detected immediately \citep[e.g.][]{Miller2015}, and others showing delayed onset of bright X-ray emission \citep[e.g.][]{Hinkle2021}. The intersection of debris streams and the circularisation process could be an important factor driving the diversity in the observational properties of many TDEs \citep{Lu2019}. 

% - development 
%     - overview of recent works

Radio emission from TDEs is rare; only $\sim10\%$ of TDEs discovered have reported radio detections. Radio observations of TDEs probe the outflowing material ejected during the stellar destruction, including any jets or wind-induced outflows, as well as their interactions with the circumnuclear medium \citep[CNM; see][for a review]{Alexander2020}. Recent radio observations of TDEs  have identified two distinct populations: relativistic, non-thermal, jetted events \citep[e.g. Swift J1644+57;][]{Bloom2011}, and the more common non-relativistic, thermal events \citep[e.g. ASASSN-14li;][]{Alexander2016,vanVelzen2016}, as well as highlighting the diverse characteristics of individual events within these populations. 

Non-thermal TDEs are thought to produce a relativistic jet, giving rise to bright radio emission with luminosities $>10^{40}$\,erg\,s$^{-1}$ \citep{Levan2011,Burrows2011,Zauderer2011,Bloom2011}. In contrast, thermal TDEs exhibit radio emission with spectral luminosities $<10^{40}$\,erg\,s$^{-1}$ that often is observed within months after the initial optical flare, and rises to a peak within a couple of years depending on the frequency \citep[e.g.][]{Alexander2016,Anderson2020,Cendes2021,Goodwin2022}. Recently, it has been suggested that delayed radio flares are common in TDEs \citep{Horesh2021,Horesh2021b,Cendes2022,Perlman2022}. However, without continuous radio coverage of the TDE lightcurve, it cannot be determined if these are "flares" or simply a slow rise to the radio peak with a structured radio lightcurve, as was the case for AT2019azh \citep{Goodwin2022,Sfaradi2022}. There are two strong cases in which there is evidence that a delayed mildly relativistic jet was produced $>500$\,d post initial disruption: ASASSN-15oi and AT2018hyz \citep{Horesh2021,Cendes2022}.

The radio emission from thermal TDEs is thought to arise from either a mildly-collimated, sub-relativistic jet \citep[e.g.][]{vanVelzen2016}, a spherical accretion-induced wind outflow \citep[e.g.][]{Alexander2016}, the unbound debris stream \citep[e.g.][]{Krolik2016}, or a spherical outflow from stream-stream collisions during the circularisation of the stellar debris \citep[e.g.][]{Lu2019}. Existing radio observations of thermal TDEs have been unable to convincingly discern the mechanism behind the non-relativistic outflows that have been observed, and new observations are crucial in identifying if there is a single mechanism behind all radio outflows from TDEs, or if it differs from system to system. 

%     - gap in research
%     - specific problem to be addressed

% -closing
%     - overview of current work
In this work we present the radio detection of AT2020opy, including three epochs of radio spectral observations of the event over 8 months. In Section \ref{sec:observations} we describe the radio observations and the data processing. In Section \ref{sec:results} we present the results and synchrotron modelling of the outflow. In Section \ref{sec:discussion} we discuss the implications of the results and provide a comparison of AT2020opy with other TDEs, and finally in Section \ref{sec:conclusion} we summarise this work and provide concluding remarks. 

\section{Observations}\label{sec:observations}
The TDE AT2020opy (ZTF20abjwvae) was first detected on 2020 July 08 by the Zwicky Transient Facility (ZTF) as a transient coincident with the nucleus of the galaxy SDSS J155625.72+232220.6 \citep{ZTF_ATATel}. The source rose slowly to a peak optical flux of $g=18.9$\,mag on 2020 August 02 and showed a featureless blue continuum in spectral observations taken with the Palomar 60in SED Machine. \textit{Swift} follow-up observations on 2020 August 09 revealed bright UV emission from the event but no associated X-ray source, motivating \citet{ZTF_ATATel} to classify the transient as a TDE. Based on the ZTF observations and optical spectral properties, \citet{Hammerstein2022} classified AT2020opy as an H+He TDE at a redshift of $z=0.159$ due to broad H$\alpha$ and H$\beta$ emission lines, as well as a complex of He II emission lines. 

\subsection{VLA}

We observed the optical position of AT2020opy on four occasions with the Karl G. Jansky Very Large Array (VLA; Proposal ID: 20A-392, PI: Van Velzen) between 2020 October 06 and 2021 June 03. Our initial observation on 2021 October 06 was taken at 8--12\,GHz to search for radio emission from the event. We discovered a point source consistent with the optical position of the galaxy with a flux density of 65$\pm$7\,$\mu$Jy at 10\,GHz. We subsequently triggered 3 epochs of radio spectral observations of the source spanning 2--18\,GHz over 8 months. The radio observations are summarised in Table~\ref{tab:observations}. 

\begin{table}
	\centering
	\caption{Dedicated radio observations of AT2020opy. $\nu$ is the central frequency of each sub-band (with bandwidth of 1\,GHz at S-band and C-band, 2\,GHz at X-band, 3\,GHz at Ku-band, and 0.856\,GHz for MeerKAT L-band), $F_{\nu}$ is the measured flux density of the source, and "Array" describes the VLA array configuration for the observations.}
	\label{tab:observations}
	\begin{tabular}{lcccr} % four columns, alignment for each
		\hline
		Date (UTC)& Array & Band & $\nu$ (GHz) & $F_{\nu}$ (uJy)\\
		\hline
		\hline
		06-Oct-2020 17:19:14 & VLA-B & X & 10 & 65$\pm$7 \\
		\hline
		15-Oct-2020 23:15:51 & VLA-B & X & 11 & 61$\pm$6\\
		 & & X & 9 & 68$\pm$10 \\
		 & & C & 4.55 & 40$\pm$13 \\
		 & & C & 5.04 & 47$\pm$16 \\
		 & & C & 6.13 & 51$\pm$63 \\
		 & & C & 7.6 & 63$\pm$12 \\
		% & & S & \\
		\hline
		17-Dec-2020 13:35:07 & VLA-A & Ku & 16.5 & 98.8$\pm$10 \\
		 &  & Ku & 13.5 & 137$\pm$10 \\
		 &  & X & 11 & 137.0$\pm$9.5 \\
		 &  & X & 9 & 134$\pm$8 \\
		 &  & C & 7.5 & 139$\pm$11 \\
		 &  & C & 6.5 & 134$\pm$15 \\
		 &  & C & 5.5 & 91$\pm$12\\
		 &  & C & 4.5 & 75$\pm$11\\
		 &  & S & 3.76 & 60$\pm$13\\
		 &  & S & 3.24 & 53$\pm$25 \\
		\hline
		03-Jun-2021 00:52:34 & VLA-C->D & X & 11 & 139$\pm$11\\
		&  & X & 9 & 198$\pm$9 \\
		 & & C & 7.5 & 175$\pm$14 \\
		 & & C & 6.5 & 241$\pm$16 \\
		 & & C & 5.5 & 256$\pm$17 \\
		 & & C & 4.5 & 252$\pm$17 \\
		 &  & S & 3.5 & 228$\pm$27\\
		 &  & S &2.5 & 251$\pm$100 \\
		 14-Aug-2021 17:42:51 & MeerKAT & L & 1.28 & 94$\pm$15 \\
		\hline
		11-May-2022 22:09:16 & MeerKAT & L & 1.28 &141$\pm18$ \\
		\hline
		23-Jun-2022 14:20:56 & uGMRT & L & 1.26 & 141$\pm$29\\
		 & uGMRT & P & 0.65 & $<261$\\
		\hline
	
	\end{tabular}
\end{table}

All VLA data were reduced in the Common Astronomy Software package \citep[CASA 5.6.3,][]{McMullin2007} following standard procedures using the VLA pipeline. For all observations, 3C 286 was used for flux density calibration, J1609+2641 was used for phase calibration for frequency ranges 2--12\,GHz (S, C, and X-band), and  J1619+2247 was used for phase calibration for 12--18\,GHz (Ku-band). To extract the source flux density, images of the target field were made using the CASA task {\sc tclean} and the flux density was measured in the image plane by fitting an elliptical Gaussian point source fixed to the size of the synthesised beam using the CASA task {\sc imfit}. We split each frequency band into 2 or 4 sub-bands depending on the bandwidth available after radio frequency interference (RFI) flagging. 

\subsection{MeerKAT}
\label{sMKTobs}

We observed AT2020opy with MeerKAT during an observing run on 2021
Aug. 14 and 2021 May 11.  We used the ``4K'' (4096-channel) wideband continuum mode
and observed with bandwidth of 856~MHz around a central frequency of
1.28~GHz, over a total time of about 3.7 hr of which $\sim$1~hr was
spent on-source for AT2020opy.

The data were reduced using the OxKAT scripts \citep{Heywood2020}.  We
used observations of 3C~286 (ICRF J133108.2+303032) to set the flux
density scale and calibrate the bandpass, and PKS J1609+2641 (ICRF
J160913.3+264129) as a secondary calibrator.  The final images were
made using the WSClean ($w$-stacking CLEAN) imager
\citep{Offringa+2014, OffringaS2017}, and had a restoring
beam of $12.9\arcsec\ \times 5.2\arcsec$ at $-20\deg$.
%and resolved into 8 layers in
%frequency.  WSClean deconvolves the 8 frequency layers together by
%fitting a polynomial in frequency to the brightness in the 8
%frequency-layers.
We obtained the final flux densities by fitting elliptical Gaussians
to the image.  Since there is a relatively nearby confusing source
with a flux density similar to that of AT2020opy, about 8\arcsec\ to
the southwest, we simultaneously fitted two elliptical Gaussians, one
for AT2020opy and one for the confusing source, along with a
zero-level to account for any constant offsets in the flux of the image.  Our value
for the flux density of AT2020opy is the flux density of the fitted
Gaussian, and the uncertainty includes both the statistical uncertainty
and a systematic one due to the uncertainty in the flux-density
bootstrapping, estimated at 5\%.

\subsection{uGMRT}

We observed AT2020opy with the upgraded Giant Metrewave Telescope (uGMRT) on 2022 June 23 at band 4 (total bandwidth of 300\,MHz with a central frequency of 0.65\,GHz) and band 5 (total bandwidth of 460\,MHz with a central frequency of 1.26\,GHz) over a total time of 3\,hr, with 51\,min on target at band 4 and 34\,min on target at band 5. Each frequency band was broken into 2048 spectral channels.
Data reduction was carried out in CASA using standard procedures including flux and bandpass calibration with 3C286 and phase calibration with ICRF J160913.3+264129. Images of the target field were created using the CASA task \texttt{tclean}. Two phase only and three phase and amplitude rounds of self-calibration were carried out on the band 4 data. As with the VLA images, the target flux density was extracted in the image plane at both bands using the CASA task \texttt{imfit}. Unfortunately no detection of the source was obtained in the band 4 observation due to a nearby bright source on the edge of the primary beam causing a high image rms. We instead report the 3$\sigma$ upper limit that was obtained at 0.65\,GHz. A 5$\sigma$ detection of the source was obtained at 1.26\,GHz (band 5) and reported in Table~\ref{tab:observations}.

% band 4:
% 2 hour observation, 51 mins on source
% chans 2048
% chan width 97.7kHz 
% total bandwidth 0.2 kHz
% central freq 650 MHz

% 2 phase only and 3 phase and amplitude selfcal, still no detection with 3 sigma upper limit of 261uJy

% band 5:
% 1 hour observation, 34 mins on source
% chan width 195.3kHz
% total bandwidth: 0.4 kHz
% center freq 1260 MHz

\subsection{Archival radio observations}

To explore the possibility of previous AGN activity in the host galaxy, we searched the radio archives for observations covering the coordinates of AT2020opy. The VLA Sky Survey \citep[VLASS,][]{Lacy2020} observed the coordinates of AT2020opy at 3\,GHz on 2020 July 16 and 2017 September 25 (35 months pre-optical flare). There was no detection of the host galaxy in either observation, with 3$\sigma$ upper limits of 490$\,\mu$Jy and 340$\,\mu$Jy respectively. The NRAO VLA Sky Survey \citep[NVSS,][]{Condon1998} also observed the coordinates of AT2020opy at 1.4\,GHz on 1995 February 28, but did not detect the host galaxy with a 3$\sigma$ upper limit of 2.1\,mJy. These observations rule out the possibility of bright ($>300\,\mu$Jy) AGN activity in the host galaxy in the past 20\,yr, but we cannot eliminate the possibility of the galaxy hosting a low luminosity AGN.

\section{Results} \label{sec:results}
\begin{figure}
	\includegraphics[width=\columnwidth]{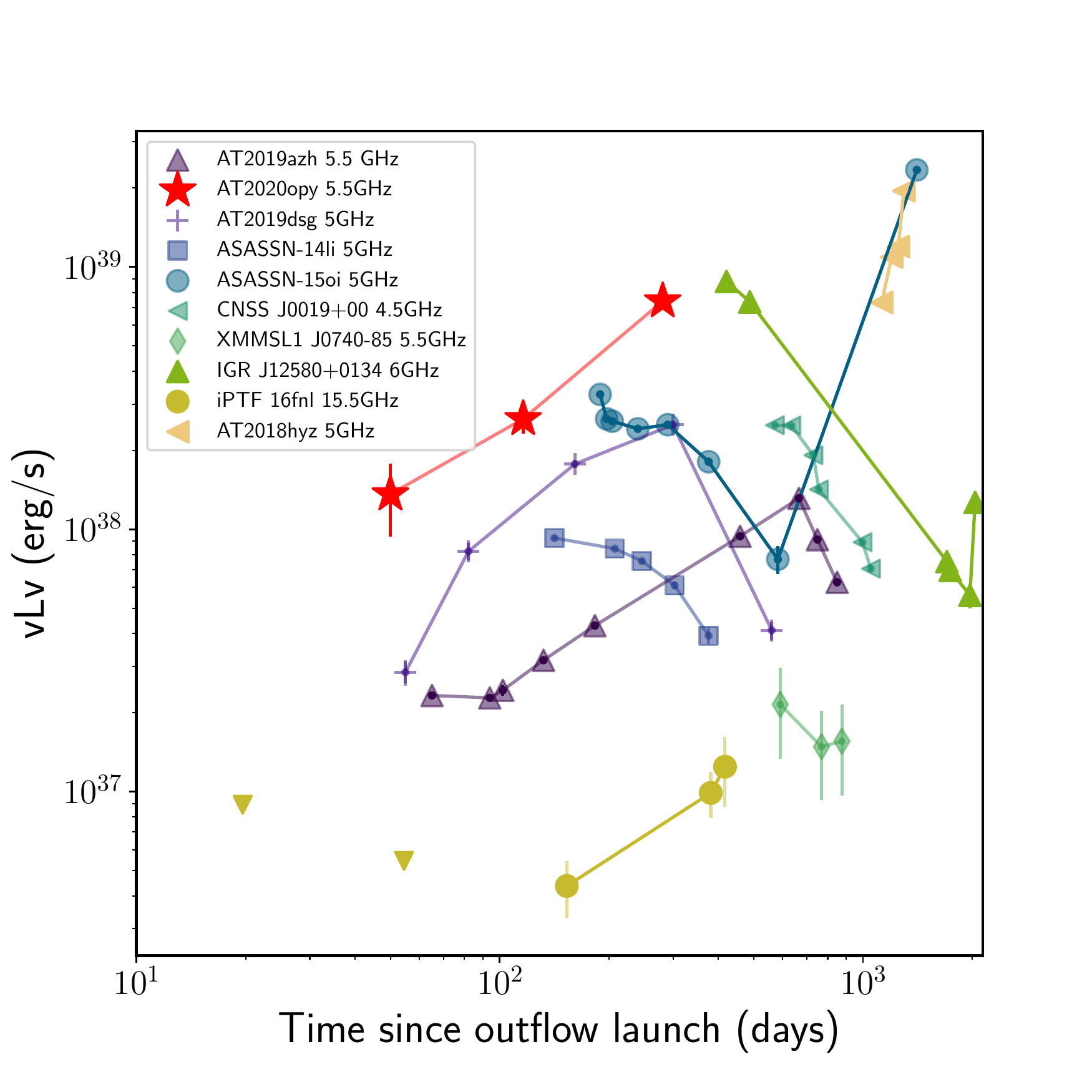}
    \caption{Radio spectral luminosity of AT2020opy at 5.5\,GHz compared to a selection of other radio-bright thermal TDEs. 
	AT2020opy is the most radio luminous TDE at early times found to date. TDE data are from AT2019azh \citet{Goodwin2022}; AT2019dsg \citet{Cendes2021}; ASASSN-14li \citet{Alexander2016}; ASASSN-15oi \citet{Horesh2021}; AT2018hyz \citet{Cendes2022}; CNSS J0019+00 \citet{Anderson2020}; XMMSL1 J0740-85  \citet{Alexander2017}; IGR J12850+0134 \citet{Perlman2022}, \citet{Lei2016}, and \citet{Nikolajuk2013}; iPTF 16fnl \citet{Horesh2021b}.}
    \label{fig:LC_comparison}
\end{figure}

The VLA radio lightcurve at 5.5\,GHz for AT2020opy compared to other radio-bright thermal TDEs is shown in Figure \ref{fig:LC_comparison} and the broadband radio spectra for each of our three epochs are plotted in Figure \ref{fig:spectra}. 

AT2020opy appears brighter than other thermal TDEs at early times relative to the outflow launch date, with luminosity $\nu L_{\nu}\approx5\times10^{38}$\,erg\,s$^{-1}$, but is not as luminous as the relativistic event Swift J1644+57 \citep[$L_{\nu}\approx2\times10^{45}$\,erg\,s$^{-1}$,][]{Zauderer2011}. The radio emission from AT2020opy is well-described by a peaked synchrotron spectrum that evolves on timescales of months, consistent with an outflow travelling through the circumnuclear medium surrounding the SMBH and accelerating electrons along magnetic field lines in the resulting shock front. 

\subsection{Synchrotron spectral fitting}

We fit the synchrotron spectra of AT2020opy using the same approach outlined in \citet{Goodwin2022}. We apply the \citet{Granot2002} model assuming the synchrotron self-absorption frequency is associated with the peak of the spectrum and $\nu_{\rm{m}} < \nu_{\rm{a}} < \nu_{\rm{c}}$, where $\nu_{\rm{m}}$ is the synchrotron minimum frequency, $\nu_{\rm{a}}$ is the synchrotron self-absorption frequency, and  $\nu_{\rm{c}}$ is the synchrotron cooling frequency. This approach enables the total spectral flux density to be modelled as a function of frequency in order to constrain the break frequencies and electron energy index, $p$. We assume no contribution to the radio emission from the host galaxy due to no previous radio detections of the host in archival observations indicating that the dominant contribution to the radio emission is the synchrotron transient component, and because earlier observations of the transient were significantly fainter than later observations.

As in \citet{Goodwin2022}, we use a Python implementation of Markov Chain Monte Carlo (MCMC), \texttt{emcee} \citep{emcee} to marginalise over the synchrotron model parameters to determine the best fit parameters and uncertainties. Due to the paucity of the data at high frequencies, we fix the synchrotron energy index to $p=2.7$ \citep[e.g.][]{Cendes2021}, but note that the derived parameters do not deviate significantly from the 1$\sigma$ uncertainty ranges if we instead choose other reasonable values, such as $p=2.5$ or $p=3$. Furthermore, $p=2.7$ is the best-fit spectral index when we allow $p$ to be a free parameter while fitting the third epoch in which the optically thin slope is best-constrained by the data. 

\begin{figure}
	\includegraphics[width=\columnwidth]{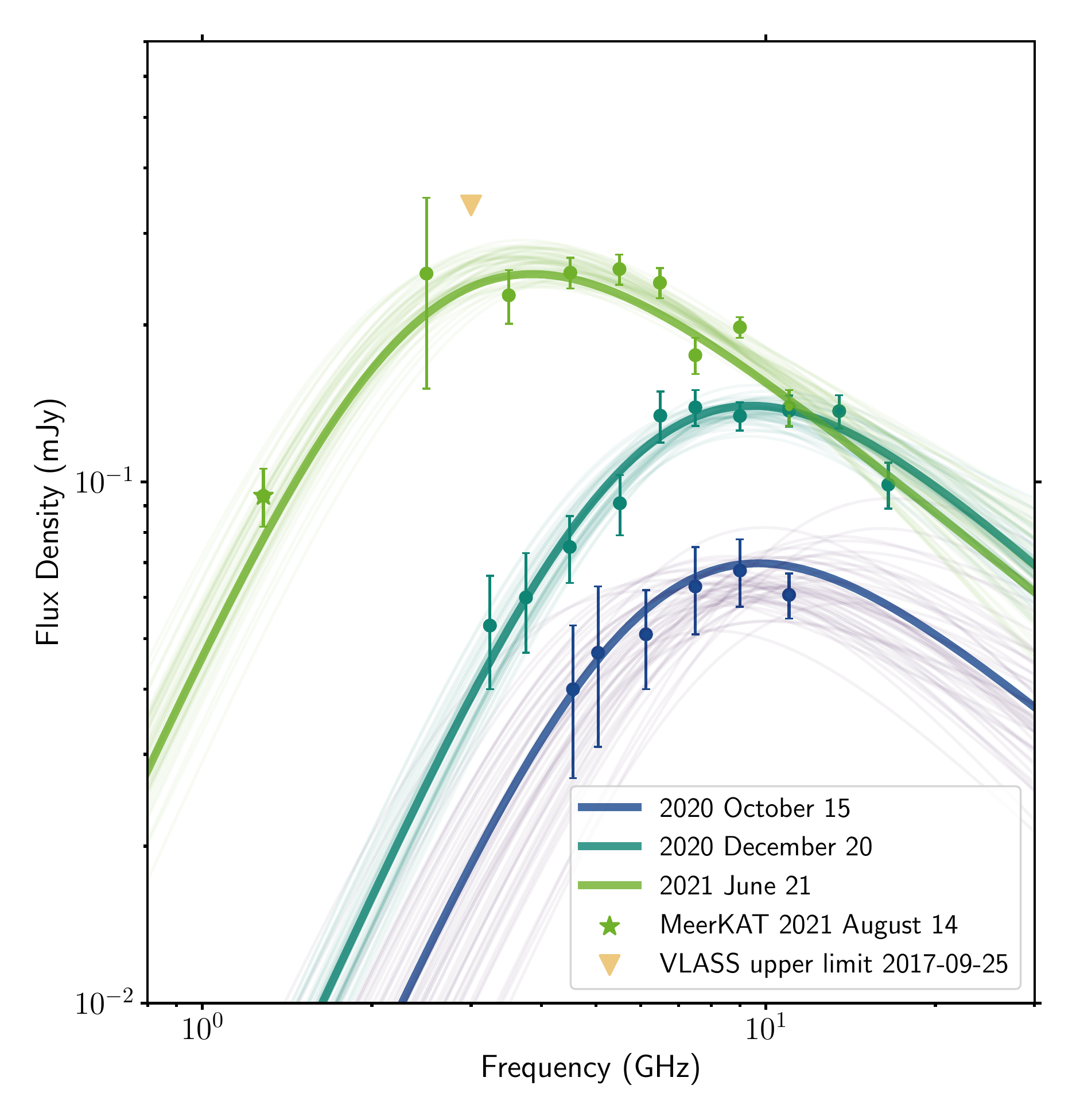}
    \caption{Radio spectra of the TDE AT2020opy obtained with the VLA (circles) and MeerKAT (star) telescopes and the median synchrotron spectra for the corresponding epochs (solid lines). 50 random samples from the MCMC fitting are plotted for each spectrum to demonstrate the uncertainty in the fits. The radio emission from AT2020opy is well-fit by an evolving synchrotron spectrum where the peak is associated with synchrotron self-absorption in each epoch.}
    \label{fig:spectra}
\end{figure}

The observed and modelled synchrotron spectra for AT2020opy are plotted in Figure \ref{fig:spectra}, and the best-fitting peak flux density and peak frequency for each epoch are listed in Table \ref{tab:spectralfits}. The synchrotron peak flux density rose consistently between the three epochs and the peak frequency decreased between the epochs. 

\subsection{Outflow modelling}

\begin{figure*}
	\includegraphics[width=\textwidth]{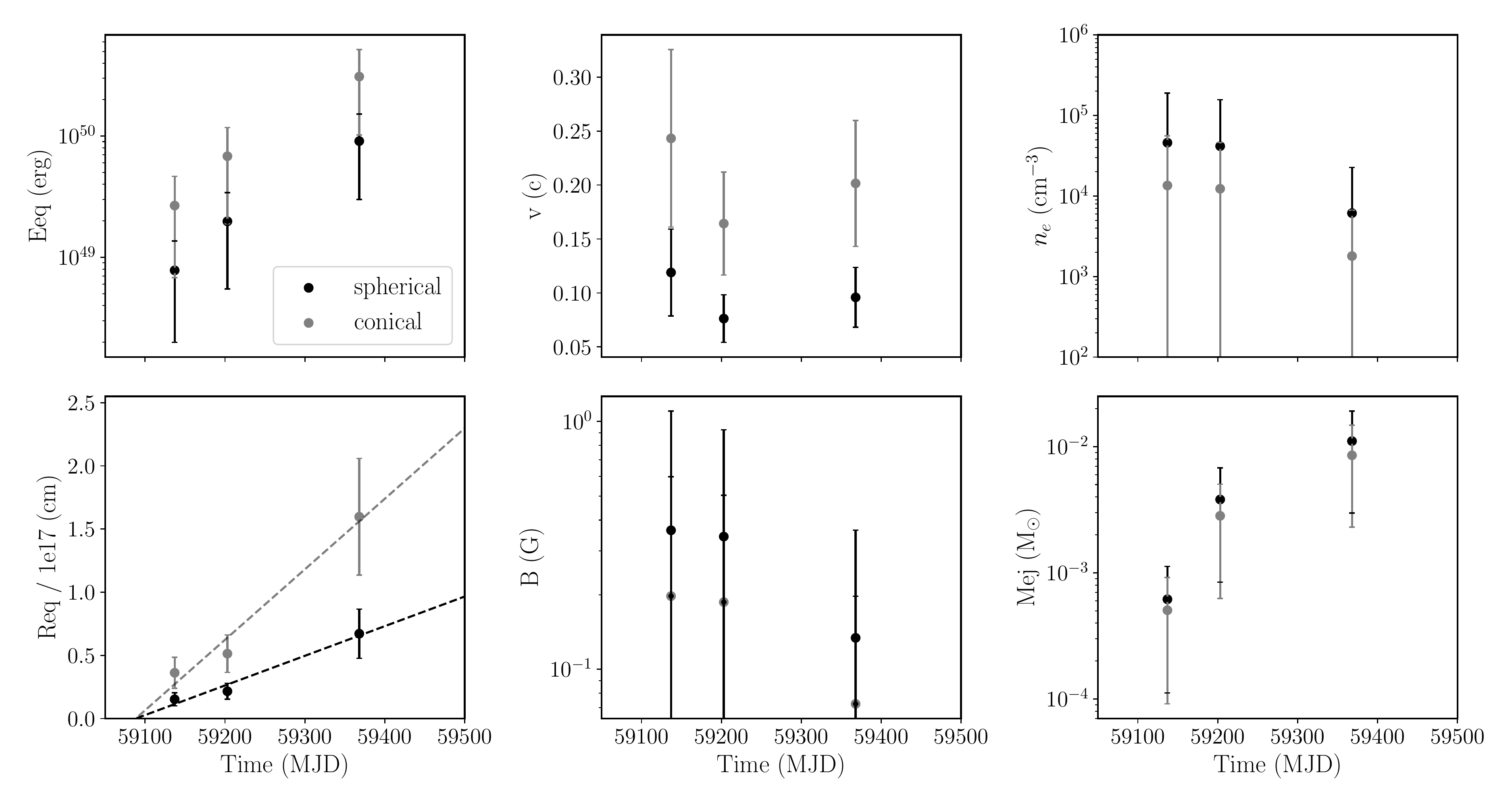}
    \caption{Inferred physical outflow properties from an equipartition analysis of three epochs of the radio synchrotron emission from AT2020opy. Black circles indicate parameters assuming a spherical homogeneous outflow, grey circles indicate parameters assuming a collimated, conical outflow. The dashed lines show a linear fit to the radius for both the spherical (black) and conical (grey) geometries, which give an estimated outflow launch date of MJD 59087$\pm$41\,d.}
    \label{fig:outflowmodel}
\end{figure*}

We model the radio outflow based on the inferred synchrotron emission properties using the same approach outlined in \citet{Goodwin2022}, in which following the model of \citet{BarniolDuran2013} we assume the ambient electrons are accelerated into a power-law distribution by the blastwave from the outflow, $N(\gamma) \propto \gamma^{-p}$, where $\gamma$ is the electron Lorentz factor. In order to estimate the outflow radius, energy, magnetic field strength, and velocity, we assume equipartition between the electron and magnetic field energy densities, which enables the derivation of an equipartition radius and energy. Once the equipartition radius and energy are obtained, we then parameterise the deviation from equipartition to derive the total energy and radius, from which other parameters can be derived.  We refer the reader to equations 4--13 of \citet{Goodwin2022} for the specific equations also used in this work. To account for different outflow mechanisms, we model two geometries of the outflow: a spherical outflow and a mildly collimated conical outflow with a half opening angle of 30\degr. We note that we model the outflow as non-relativistic (bulk Lorentz factor $\Gamma=1$), as a relativistic outflow is only possible for very small ($\lesssim0.1$\,deg) opening angles.  The estimated physical outflow properties for AT2020opy are plotted in Figure \ref{fig:outflowmodel} and listed in Table \ref{tab:spectralfits} for each of these geometries.

\begin{table*}
\label{tab:spectralfits}
    \centering
        \caption{Synchrotron spectral fitting parameters and outflow model predictions for AT2020opy. \textbf{Fix values in table}}
	\begin{tabular}{lccccccccr}
		\hline
		 & $\delta t$ (d) & $\nu_{\mathrm{peak}}$ (GHz) & $F_{\rm peak}$ (mJy) & log$_{10}$\,$R$ (cm) & log$_{10}$\,$E$ (erg) & $\beta$ & log$_{10}$\,$B$ (G) & log$_{10}$\,$n_e$ (cm$^{-3}$) \\
		\hline
		\hline
        & 50 & $9.6\pm1.2$ & $0.070\pm0.007$ & $16.2\pm0.1$ & $48.9\pm0.3$ & $0.12\pm0.04$ & $-0.4\pm0.9$ & $4.7\pm1.3$\\
       Spherical & 116 & $9.4\pm0.5$ & $0.141\pm0.006$ & $16.3\pm0.1$ & $49.3\pm0.3$ & $0.08\pm0.02$ & $-0.5\pm0.7$ & $4.6\pm1.2$ \\
        & 281 & $4.2\pm0.4$ & $0.28\pm0.02$ &  $16.8\pm0.1$ & $50.0\pm0.3$ & $0.10\pm0.03$ & $-0.9\pm0.7$ & $3.8\pm1.2$ \\
        
        \hline
        & 50 & $9.6\pm1.2$ & $0.070\pm0.007$ & $16.6\pm0.1$ & $49.4\pm0.3$ & $0.24\pm0.08$ & $-0.7\pm0.9$ & $4.1\pm1.3$ \\
       Conical & 116 & $9.4\pm0.5$ & $0.141\pm0.006$ & $16.7\pm0.1$ & $49.8\pm0.3$ & $0.16\pm0.05$ & $-0.7\pm0.7$ & $4.0\pm1.2$ \\
        & 281 & $4.2\pm0.4$ & $0.28\pm0.02$ &  $17.2\pm0.1$ & $50.5\pm0.3$ & $0.2\pm0.06$ & $-1.1\pm0.7$ & $3.3\pm1.2$ \\
        \hline
    \end{tabular}

\textit{Note:} $\delta t$ is reported with reference to $t_0$, the estimated outflow launch date of MJD 59087. 
\end{table*}

The radius increased approximately linearly with time, indicating approximately constant velocity of the outflow. A simple linear fit to the radius (Figure \ref{fig:outflowmodel}) gives an outflow launch date of MJD=59087$\pm$41\,d or MJD=59088$\pm$43\,d  for spherical and conical geometries, respectively; 50\,d after the optical flare was first observed. This predicted outflow launch date is coincident with the optical peak, on MJD 59070, and is also coincident within 2-$\sigma$ of the initial optical flare on MJD=59038. 

The energy of the outflow increased approximately linearly with time, as is expected for an increasing synchrotron peak flux density, which could be indicative of constant energy injection into the outflow. The velocity and magnetic field strength remained approximately constant at 0.1\,$c$ (0.2\,$c$) and 0.28\,G (0.15\,G) for the spherical (conical) geometries, with no sign of relativistic motion of the outflow. We note that the inferred radius, energy, and ambient density are consistent with having remained constant between the first and second epochs of observations at 50 and 116\,d post outflow launch. However, due to the paucity of data in the first epoch and the large uncertainties in the resulting spectral fits, we deduce that it is more likely that the outflow was evolving between these two epochs, as there is signficant evolution between the first and last epochs. 

\section{Discussion}\label{sec:discussion}
%summarise properties
The radio properties of the TDE AT2020opy indicate a non-relativistic outflow was launched at the time of or just after the initial optical flare. We deduce that the outflow has an approximately constant velocity with $\beta\approx0.1\,c$ and energy $\sim10^{48}$\,erg for radii $\sim10^{16}$\,cm. Between 2020 October and 2021 June the radio emission from AT2020opy was increasing in peak flux density and the peak frequency of the synchrotron spectrum was decreasing, consistent with a constant velocity outflow moving through the CNM and sweeping up material.

%combine interpretation with optical properties of the host
\citet{Hammerstein2022} analysed the optical spectra of AT2020opy and classified the event as a TDE with broad H$\alpha$ and H$\beta$ emission lines as well as a complex of He II emission lines (H+He spectral class) and a structured optical lightcurve with some flaring activity. In an analysis of 30 TDEs, \citet{Hammerstein2022} found some evidence that TDEs with structured lightcurves tend to occur in galaxies with lower total mass, and thus could occur around lower mass SMBHs. We note that this finding is strongly dependent on an assumption of a relationship between total stellar mass in the host galaxy and SMBH mass. \citet{Hammerstein2022} propose that the structured flaring activity seen in the lightcurves of these TDEs could be due to longer circularisation times of the lower mass SMBHs. A longer circularisation time of the disk for AT2020opy could explain the lack of early X-ray emission from the event \citep{ZTF_ATATel}, as well as a radio outflow that was launched after the initial optical flare. 

\subsection{The outflow mechanism}

The radio measurements that enable determinations about the physical properties of the outflow produced in AT2020opy enable some discrimination between current models of non-relativistic outflows in TDEs. Firstly, the data indicate that the outflow in this event is likely to have been launched approximately 50\,d after the initial optical flare, however we cannot rule out a contemporaneous launch of the outflow at a high degree of confidence (we infer it was launched at least 8\,d after to 1$\sigma$). In comparison, the radio outflows observed for the thermal TDEs AT2019azh and ASASSN-14li were inferred to have been launched at the time of the initial optical flare \citep{Goodwin2022,Alexander2016,vanVelzen2016}. 
Secondly, the observed velocity of the outflow is approximately constant (under the assumption of ballistic motion) and the radio emission cannot be explained by a relativistic outflow unless the jet has an unphysically small opening angle.
\citet{Hammerstein2022} found some evidence that TDEs with structured lightcurves occur in lower mass host galaxies, which could translate to lower mass central black holes with longer circularisation times, making stream-stream collisions more important. A long circularisation time of the stellar debris for AT2020opy could explain a delayed onset of the radio outflow if the outflow was produced by either a disk wind or debris collisions. Finally, the lack of X-ray emission from the event at early times \citep{ZTF_ATATel} could either be due to intrinsically low X-ray activity due to the accretion disk taking a long time to form, making the super-Eddington accretion induced wind outflow scenario unlikely, or due to the large distance to the source.

We thus conclude that the outflow from AT2020opy is more likely to be explained by a spherical outflow from stream-stream collisions of the circularising stellar debris \citep{Lu2019} than by an accretion induced wind outflow from accretion onto the SMBH \citep[e.g.][]{Alexander2016}. We deduce that the unbound debris stream is unlikely to explain the radio properties of this outflow due to the low mass predicted in the outflow ($\lesssim 10^{-2}M_{\odot}$) and predicted small opening angle of the unbound debris stream \citep[e.g.][]{Guillochon2014}. 

Under the assumption that the outflow was produced by a collision induced outflow (CIO) or accretion induced wind, the increasing energy with constant velocity of the radio outflow is consistent with a single injection of energy into the outflow that is sweeping up material from the CNM \citep{Lu2019}. Under such an assumption, we can approximately infer the deceleration time of the outflow using the model from \citet{Lu2019}, where for $k=2$, the deceleration radius is given by

\begin{equation}
    r_{\mathrm{dec}}= \frac{1}{\Omega} \frac{2 E_{\rm k}}{N m_{\rm p} v_0^2}
\end{equation}
where the outflow is assumed to be a thin shell covering solid angle $\Omega\sim2\pi$, $E_{\rm k}$ is the kinetic energy in the outflow, $m_{\rm p}$ is the proton mass, $N$ is the electron density, and $v_0$ is the average outflow velocity. 

\begin{figure}
    \centering
    \includegraphics[width=\columnwidth]{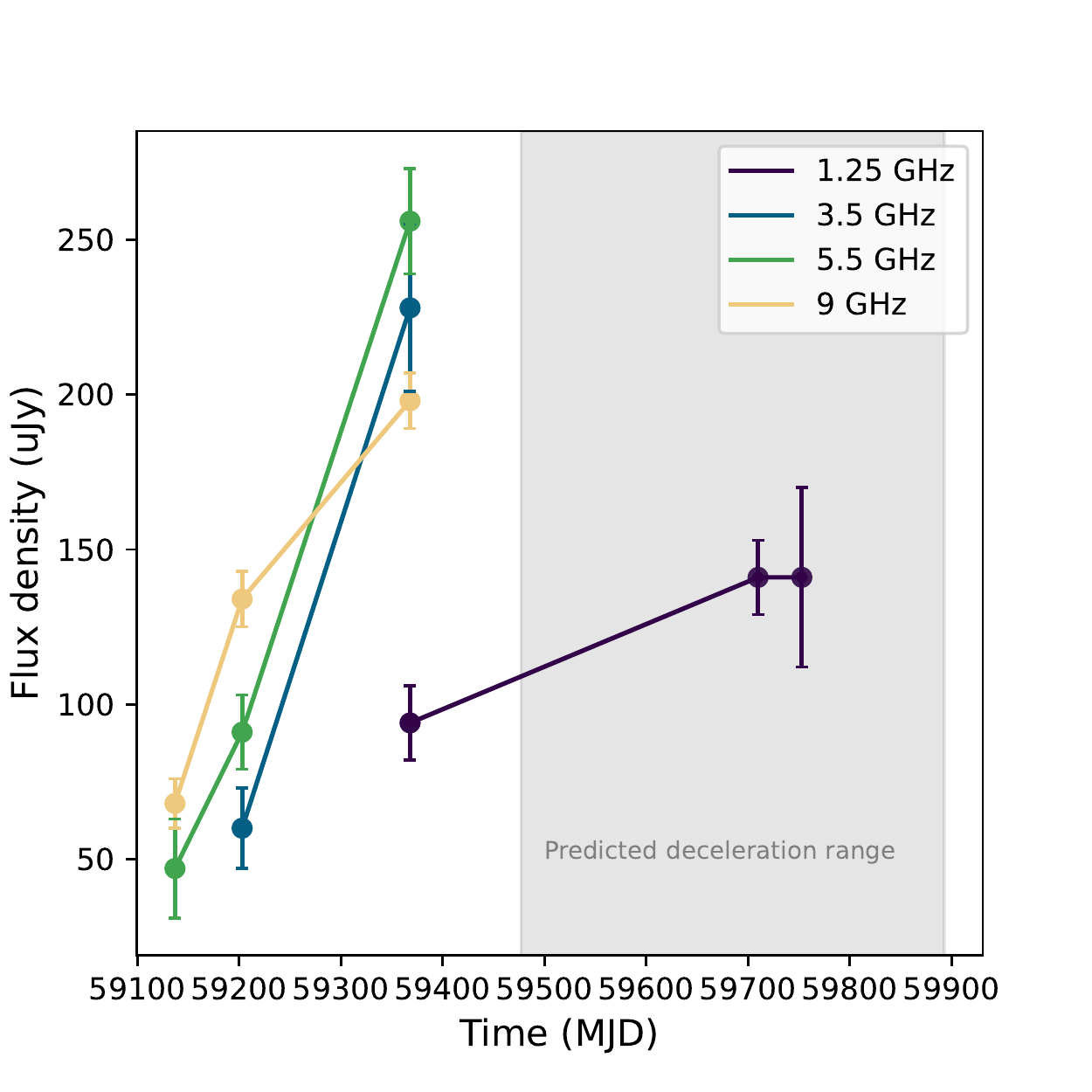}
    \caption{Radio lightcurve of AT2020opy at 1.25, 3.5, 5.5, and 9\,GHz. During the first $\approx300$\,d post-disruption the radio flux density rose at all frequencies. At 623--666\,d post-disruption the radio flux density appears to have plateaued at 1.25\,GHz, indicating that the outflow may have begun to decelerate.}
    \label{fig:lightcurve}
\end{figure}

For $E_{\rm k}\sim5\times10^{51}$ ($10^{52})$\,erg, $v_0=0.1\,c$, and $N=10^{3}$\,cm$^{-3}$, we infer $r_{\mathrm{dec}}\approx10^{17}$\,cm ($2\times10^{17}$\,cm), and thus $t_{\mathrm{dec}}\approx$390\,d (780\,d).  At 281\,d we observed the radio outflow to still be expanding and increasing in luminosity/energy(Figure \ref{fig:outflowmodel}), and at 623--666\,d we observed the radio emission to still be increasing/constant in luminosity at 1.25\,GHz (Figure \ref{fig:lightcurve}). We thus predict that the outflow could still be increasing in luminosity until 2022 November, depending on the kinetic energy available in the outflow. Further radio observations of the event would help constrain the deceleration radius/time of the outflow and enable further discrimination between outflow models. Importantly, we note that for this kind of outflow, the radio emission can continue to increase for up to years after the initial event, depending on the energy available in the outflow and the density of the CNM.  For AT2020opy, the observed radio emission is more luminous likely due to a denser CNM, and thus may peak earlier than other thermal TDEs. 

\subsection{Comparison with other TDEs}

\begin{figure*}
	\includegraphics[width=0.45\textwidth]{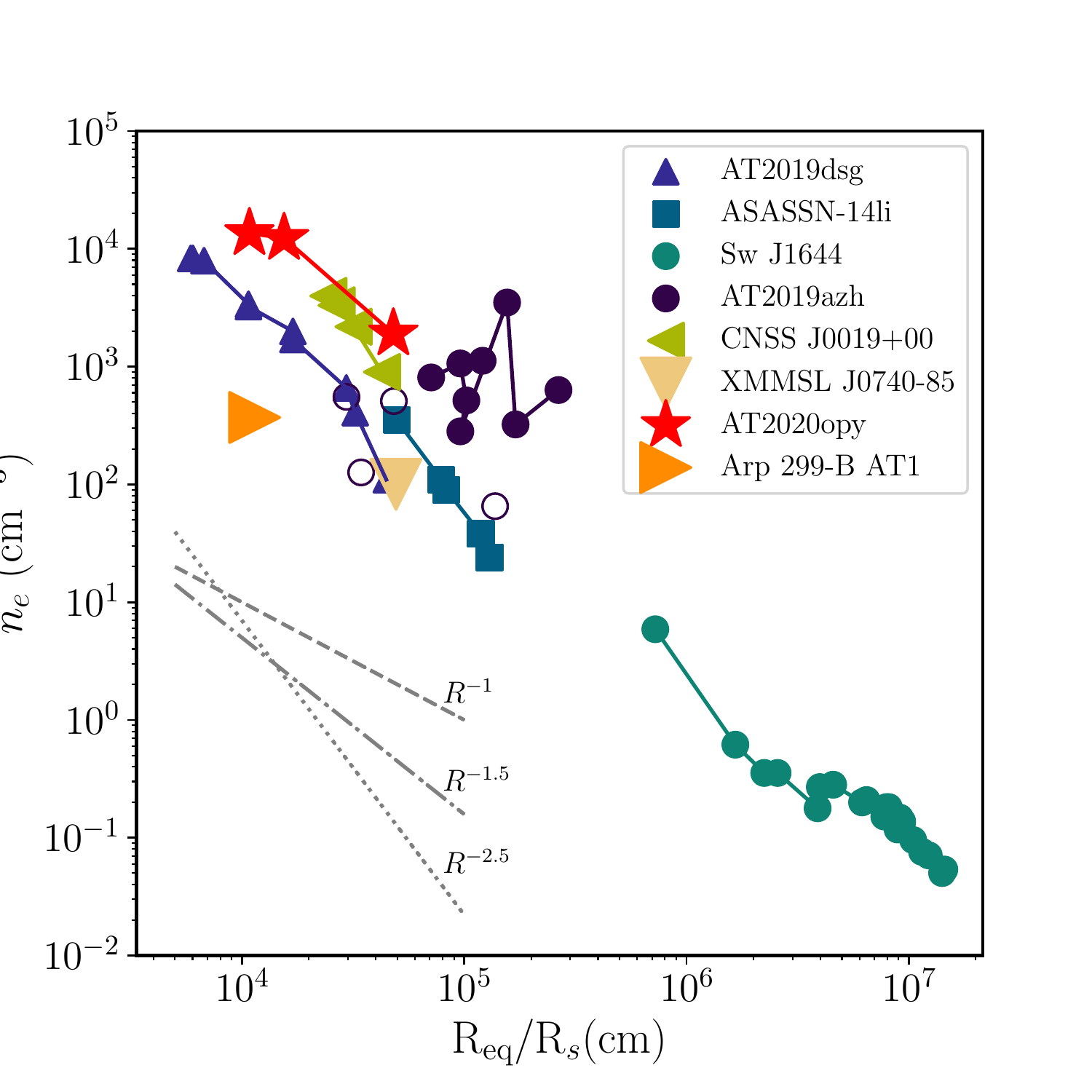}
	\includegraphics[width=0.45\textwidth]{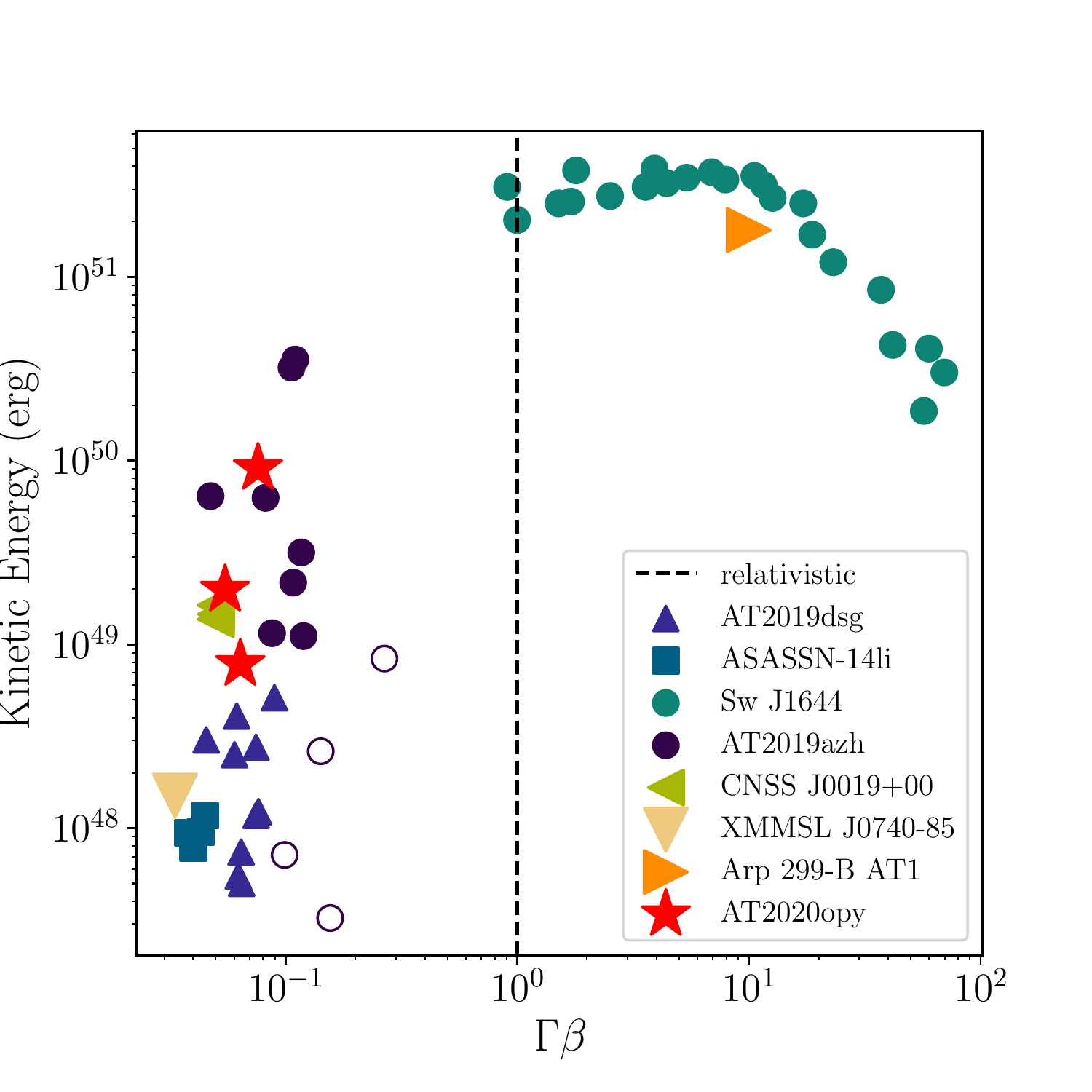}
    \caption{\textit{Left:} The scaled radius and ambient density for a selection of TDEs. \textit{Right:} The kinetic energy and outflow velocity for a selection of TDEs. TDE data and assumed SMBH masses are from \citet{Cendes2021,Stein2021} (AT2019dsg, $M_{\mathrm{BH}}=5\times10^6$\,$M_{\mathrm{\odot}}$), \citet{Alexander2016} (ASASSN-14li, $M_{\mathrm{BH}}=1\times10^6$\,$M_{\mathrm{\odot}}$), \citet{Eftekhari2018} (Sw J1644+57, $M_{\mathrm{BH}}=1\times10^6$\,$M_{\mathrm{\odot}}$), \citet{Anderson2020} (CNSS J0019+00, $M_{\mathrm{BH}}=1\times10^7$\,$M_{\mathrm{\odot}}$), \citet{Mattila2018} (Arp 299-B AT1, $M_{\mathrm{BH}}=2\times10^7$\,$M_{\mathrm{\odot}}$), \citet{Alexander2017} (XMMSL1 J0740-85, $M_{\mathrm{BH}}=3.5\times10^6$\,$M_{\mathrm{\odot}}$), and \citet{Goodwin2022} (AT2019azh, $M_{\mathrm{BH}}=3\times10^6$\,$M_{\mathrm{\odot}}$). For AT2020opy we assume $M_{\mathrm{BH}}=1.12\times10^7$\,$M_{\mathrm{\odot}}$ \citep{Hammerstein2022}. $R_s$ is the Schwarzschild radius of the black hole and R$_{\mathrm{eq}}$ is the predicted equipartition radius of the outflow. AT2020opy fits well into the population of thermal TDEs, with a slightly higher density at radii closer to the SMBH.}
    \label{fig:comparison}
\end{figure*}

A comparison of the inferred outflow properties for AT2020opy with other radio-bright TDEs is shown in Figure \ref{fig:comparison}. AT2020opy clearly fits into the population of non-relativistic events in terms of energy, velocity, and radius from the central black hole. 

The ambient density is approximately proportional to $n\propto R^{-1.5}$--$R^{-2.5}$ for AT2020opy, similar to other thermal TDEs. The CNM density of AT2020opy appears to be $\sim30\%$ denser than any of the other TDEs observed to date, which could explain the higher luminosity (and higher distance) of the radio emission from the event. Interestingly, AT2020opy is also the most distant of the thermal TDEs with detected radio emission reported, implying that for more energetic events radio emission may be observed from further away. The inferred CNM density and radio luminosity of AT2020opy further confirm that for galaxies with higher CNM densities, outflow emission rises more quickly and is brighter than in galaxies with less-dense CNMs \citep{Lu2019}. 

The delay relative to the optical flare in radio emission observed from AT2020opy was not large, similar to the two other thermal TDEs that have been observed in the radio-rise phase AT2019azh \citep{Goodwin2022} and AT2019dsg \citep{Stein2021,Cendes2021}, in contrast to the late-time radio flare that was observed from ASASSN-15oi \citep{Horesh2021}. Our modelling constrains the onset of the radio outflow in AT2020opy to be consistent with the time of or just after the optical flare was observed.

\section{Conclusions} \label{sec:conclusion}
We followed the radio evolution of the tidal disruption event AT2020opy for 20 months with the VLA, MeerKAT, and uGMRT radio telescopes. Based on modelling of the synchrotron emission observed, we find that the radio emission is likely due to a non-relativistic outflow, which could take the form of a spherical wind, collision induced outflow or a mildly collimated jet. We note that based on modelling of the evolution of the radius, we find the outflow was launched after or around the time that the initial optical flare was observed. Through synchrotron spectral modelling of the radio emission, we deduce that the circumnuclear medium of the host galaxy is denser than inferred for other TDE hosts, which causes brighter, quickly rising radio emission from the outflow.

Follow-up observations of this event are encouraged to continue to observe the long-term decay of the radio emission, which we predict will reach a peak luminosity at 390-780\,d post-optical flare (up to 2022 November). 

\section*{Acknowledgements}

The authors thank K. Alexander, N. Blagorodnova, P. Woudt, M. Bottcher, R. Fender, J. Bright, and S. Kulkarni for their contributions to the observing proposals that were instrumental to this work. This work was supported by the Australian government through the Australian Research
Council's Discovery Projects funding scheme (DP200102471). A.H. is grateful for the support by the I-Core Program of the Planning and Budgeting Committee and the Israel Science Foundation, and support by ISF grant 647/18. A.H. is grateful for support by the Zelman Cowen Academic Initiatives. GRS is supported by NSERC Discovery Grants RGPIN-2016-06569 and RGPIN-2021-0400.
The National Radio Astronomy Observatory is a facility of the National Science Foundation operated under cooperative agreement by Associated Universities, Inc. The MeerKAT telescope is operated by the South African Radio Astronomy Observatory, which is a facility
of the National Research Foundation, an agency of the Department of Science and Innovation. We thank the staff of the GMRT that made these observations possible. GMRT is run by the National Centre for Radio Astrophysics of the Tata Institute of Fundamental Research.

%%%%%%%%%%%%%%%%%%%%%%%%%%%%%%%%%%%%%%%%%%%%%%%%%%
\section*{Data Availability}
The spectral fitting and equipartition modelling software used in this work is publicly available on Github at \url{https://github.com/adellej/tde_spectra_fit}.

\section*{Software}
This research made use of Matplotlib, a community-developed \texttt{Python} package \citep{Hunter2007}, NASA's Astrophysics Data System Bibliographic Services, the Common Astronomy Software Application package \texttt{CASA} \citep{McMullin2007}, The Cube Analysis and Rendering Tool for Astronomy \citep[CARTA][]{Comrie2021} and the \texttt{Python} packages cmasher \citep{cmasher}, and emcee \citep{emcee}.



\bibliographystyle{mnras}
\bibliography{bibfile} % if your bibtex file is called example.bib

%\begin{thebibliography}{99}
%\end{thebibliography}

%%%%%%%%%%%%%%%%%%%%%%%%%%%%%%%%%%%%%%%%%%%%%%%%%%

%%%%%%%%%%%%%%%%% APPENDICES %%%%%%%%%%%%%%%%%%%%%

% \appendix

% \section{Some extra material}

%%%%%%%%%%%%%%%%%%%%%%%%%%%%%%%%%%%%%%%%%%%%%%%%%%

% Don't change these lines
\bsp	% typesetting comment
\label{lastpage}
\end{document}